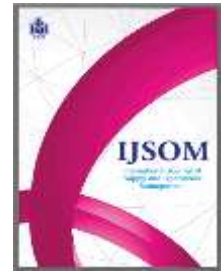

# Modelling the level of adoption of analytical tools; An implementation of multi-criteria evidential reasoning

Igor Barahona[a*], Judith Cavazos[b] and Jian-Bo Yang[c]

[a] Chapingo Autonomous University (UACh) - Cátedras CONACYT, México
[b] Popular Autonomous University of Puebla State. Puebla, México
[c] The University of Manchester. Manchester, UK

**Abstract**

In the future, competitive advantages will be given to organisations that can extract valuable information from massive data and make better decisions. In most cases, this data comes from multiple sources. Therefore, the challenge is to aggregate them into a common framework in order to make them meaningful and useful. This paper will first review the most important multi-criteria decision analysis methods (MCDA) existing in current literature. We will offer a novel, practical and consistent methodology based on a type of MCDA, to aggregate data from two different sources into a common framework. Two datasets that are different in nature but related to the same topic are aggregated to a common scale by implementing a set of transformation rules. This allows us to generate appropriate evidence for assessing and finally prioritising the level of adoption of analytical tools in four types of companies. A numerical example is provided to clarify the form for implementing this methodology. A six-step process is offered as a guideline to assist engineers, researchers or practitioners interested in replicating this methodology in any situation where there is a need to aggregate and transform multiple source data.

## 1. Introduction

The complexity of today's economic environment with market globalisations, the emergence of more powerful computers, intricate Internet-based systems, and the proliferation of real-time communication channels is transforming the way in which organisations make decisions. The first immediate result of these changes is the accumulation of massive amounts of data. According to Gantz & Reinsel (2012), data accumulated in the years 2005 to 2020 will grow by a factor of 300;

* Corresponding email address: igorbarahona@hotmail.com

that is, from 130 exabytes to 40,000 exabytes, or 40 trillion gigabytes. In terms of composition, around 68% of all information worldwide will be created and used by consumers in the form of activities such as watching digital TV, interacting in social networks, or sending images and videos, among others. Private organisations will own nearly 80% of the data in the "*digital universe*" and will be required to deal with issues as security, privacy, copyright and regulatory compliance.

In view of this exponential growth in data, it is clear that organisations will have to respond to these changes. It is a fact that traditional decision making approaches, based mostly on intuitive judgements and past experience, are gradually becoming inadequate guides for dealing with this increasingly complexity. The challenge, then, is to find new approaches for extracting relevant information from the enormous amounts of data available, and making more accurate decisions. In today's globalised markets, the ability to analyse data and create value in order to successfully response to the expectations of customers, suppliers, staff, shareholders and society will represent a competitive advantage. The emergence of the concept of evidence-based management (EBMgt) in 2006 made this tendency clear. According to Rousseau (2006), EBMgt is defined as the discipline of making the most accurate decisions through the application of science and research-principles, which is possible only when the values are credible, the evidence is clear and findings are interpretable by all stakeholders. A second movement introduced in response to the mentioned tendencies is the concept of predictive analytics. This basically deals with extracting valuable information from data in order to predict trends, behaviours and patterns. The main concept behind predictive analytics relies on establishing relations between explanatory and predicted variables ("Predictive analytics", 2014). Here, only two movements are discussed in order to illustrate what some experts and practitioners are doing as a response to the need to take advantage of "*big data*". An extensive discussion about these changes and tendencies can be found in Davenport, Harris & Morison (2010), Lynch (2008), Scott, A. J. (2012) and Anderson-Cook et al (2012). The discussion will centre in later lines on how real-world data is obtained in order to validate what we found in our review of literature. Our research objectives will be introduced at the end of this section.

The relevance of investigating how organisations can improve their analytical capabilities and obtain more benefits from data available is clear. Barahona & Riba (2012) discussed a five-level scale to measure the level of adoption of analytical tools, which was later applied to a sample of 255 organisations. An analysis of this data allowed us to formulate guidelines that would help companies to improve their analytical capabilities. The research was later complemented by interviews with managers, consultants, academics and practitioners. A total of 10 interviews were carried out, and the results allowed us to propose a three-level scale. Based on these two sources of data with different scales, the challenge is to provide a generic framework that will allow us to obtain unique and relevant conclusions, while preventing a loss of information. A novel structure is required in order to deal with this problem, as stated in the following objectives:

- Based on the principles set by Yang et al (2011), investigate the scales from questionnaires and interviews so that these can be aggregated into a unique framework.
- Apply the evidential reasoning approach to calculate the overall performance of the level of adoption of analytical tools for four types of companies.
- Offer a numerical example as a tool for replication to researchers and practitioners interested in tackling the problem of multiple-source data.

This paper is divided into five parts. The following section will review literature related to multi-criteria decision analysis methodologies. Section three gives an explanation of the methodology implemented. A numerical example is provided in section four, with our conclusions presented in the last section.

## 2. Literature review

According to Belton & Stewart (2002) the term Multi-Criteria Decision Analysis (MCDA) encompasses several quantitative approaches which "*seek to take explicit account of multiple criteria in helping individuals or groups explore decisions that matter*". Mendoza & Martins (2006) identify three formal dimensions that are present in all of these approaches: 1) All MCDA methods are based on a formal approach; 2) they are invariably formed of multiple criteria; and 3) decisions are made either by individuals or groups. Pohekar & Ramachandran (2004) emphasise some features present in all MCDA methods: first, they help to improve the quality of decisions made by making them more explicit, rational and efficient. Similarly, they increase the level of understanding of complex problems, encourage the active participation of the individuals involved, stimulate decisions made collectively and therefore improve team work skills.

Literature contains several ways to classify different MCDA methods. Mendoza & Martins (2006) explained a classification that makes a distinction between multi-objective decision making (MODM) and multi-attribute decision making (MADM). The main difference between these two families of methods is basically the number of alternatives. Hayashi (2000) suggests that MADM methods are appropriate for decisions on discrete alternatives while MODM are more suitable for tackling multi-attribute problems when infinite numbers of solutions are given by a set of previously defined constraints. A deeper discussion about the differences between multi-objective and multiple-attribute is provided in Malczewski (1999), Tzeng & Huang (2011), Pukkala (2002), Belton & Stewart (2002) and Mendoza &Prabhu (2002). These are summarised *in Table 1*.

Among the most widely used methods is the *Weighted sum method* (WSM). According to Belton & Stewart (2002) WSM is suitable for tackling single dimensional problems. For instance, consider a scenario when the best alternative will be selected from *M* different alternatives and *N* criteria. The overall performance of each alternative is equal to the sum of its products; that is: $\sum_{i=1}^{N} m_i n_i$. As the complexity of the problem increases, the suitability of the method decreases.

**Table1.** Differences between MODM and MADM methods (Adapted from Malczewski, 1999)

| Criteria for comparison | MODM | MADM |
|---|---|---|
| Now criteria are defined. | Objectives | Attributes |
| How objectives are defined. | Explicitly | Implicitly |
| How attributes are defined | Implicitly | Explicitly |
| How constrains are defined. | Explicitly | Implicitly |
| How alternatives are defined. | Implicitly | Explicitly |
| Number of alternatives | Infinite (Very large) | Finite (Small) |
| Decision maker's control | Very high | Limited |
| Decision modelling paradigm. | Design / Search | Evaluation / Choice |

Another widely known MCDA method is the *analytical hierarchy process* (AHP) proposed by Saaty (1980). This method compares pairs of alternatives to assess their relative importance in order to reach a given criterion. In terms of Pohekar & Ramachandran (2004), this is done by breaking the problem down into a hierarchy with the goal at the top, criteria and sub-criteria at levels and sublevels, and alternative decisions at the bottom level of the hierarchy. Comparisons are made by applying verbal terms that assign weights to each alternative. In this way, the final composite vector of weights is obtained, which expresses the ranked alternatives. A detailed explanation of the AHP methodology and its applications for different disciplines can be found in Saaty (1980), Saaty (1990) and Saaty (1991).

The ELECTRE, (ELimination Et ChoixTraduisant la REalité, for the French acronym) proposed by Roy (1968), is another MCDA method. Mousseau, Slowinski & Zielniewicz (2000) proposed a two step approach to assign alternatives to pre-defined criteria: first, the construction of an outranking relation *S* which characterises how alternatives compare to the limits of categories; and second, the exploitation of the relation *S* in order to assign each alternative to a specific criterion. In this way, alternative "*a*" is assigned as a result of the comparison of "*a*" with profiles defined by the limits of the categories. The criteria in this method have two distinct sets of parameters: the importance coefficients and the veto thresholds. Mousseau & Slowinski (1998) and Mousseau, Slowinski & Zielniewicz (2000) provide a detailed explanation of how specialised software can be utilised for this application (See http://www.decision-deck.org/diviz/download.html).

The fourth approach presented here is PROMETHEE (Preference Ranking organisation Method for Enrichment Evaluations). According to Brans, Vincke & Mareschal (1986), the method is based on mutual comparisons of pairs of alternatives with respect to each selected criterion. A preference function $f(a, b)$ is introduced in order to rank the investigated alternatives. That is, the criteria functions are used to assess alternatives $a$ and $b$. For instance, if the result for the given function is $f(a) > f(b)$, then we can say that alternative $a$ is better than $b$. Based on the pairwise comparisons, the decision maker assigns a preference to each alternative, in a value from 0 to 1. Louviere, Hensher & Swait (2000) and Brans, Vincke & Mareschal (1986) offer a deeper discussion on this method.

The TOPSIS method (Technique for Order Preference by Similarity to Ideal Solution) was introduced by Yoon & Hwang (1995) as an alternative methodology to ELECTRE. This method is based on the premise that the selected alternative should be the shortest geometrical distance from the negative ideal solution. One assumption used in the model is that each criterion is evenly distributed in the geometrical space and thus the order of preference of the alternatives is given by comparing Euclidean distances among alternatives. First, a decision matrix with $M$ alternatives and $N$ criteria is formulated. Then the normalised decision matrix is calculated, giving us the ideal and non-negative solutions. Like other methods, it searches for the maximum value and the minimum values amongst the alternative cost criteria. Finally, the calculations, separations and measures are made in order to get the relative closeness to the ideal solution. According to Pohekar & Ramachandran (2004), the best alternative is the one that is the shortest distance to the ideal solution and the longest distance from the negative ideal solution.

The evidential reasoning (ER) approach is a generic *evidence-based* type of multi-criteria decision analysis (MCDA). According to Yang & Singh (1994), the evidential reasoning approach is different from conventional MCDA methods in that it uses *evidence-based* reasoning to reach a decision. One of the most important contributions of this method is its capacity to describe a scenario by using belief structures or belief decision matrices, which is used to assess each alternative through vectors of paired elements. The ER approach aggregates attributes through a non-linear process that is given by the weights of criteria, and assesses the form of each criterion. This characteristic, available in ER, is not available in other MCDA methods.

Considering that the previously mentioned features of ER are not present in other MCDA methods, this is a suitable tool for investigating the level of adoption of analytical tools. The problem of aggregating data from different sources by applying the ER approach is discussed in the following sections.

## 3. Methodology

A general notion defines the concept of prioritising as assigning a higher value to some things over other things. Yang et. al (2011) define prioritising as ranking the alternatives for either a given individual criteria or for the overall criterion. For example, one simple approach for ranking the level of adoption of analytical tools is to quantify each value of the scale into a certain fixed value, calculate its mean, and rank different alternatives based on their mean values. The problem is that this approach can only produce a narrower sense of the data, while any richer information contained on it is lost. Yang et. al (2011) propose a solution to this problem, which is to utilise a generic framework. This method does not require the assessment values to be quantified to fixed values; instead, it allows to them to take any values that suit their qualitative definitions and meanings. The implementation of this approach is explained in following paragraphs.

Figure 1 introduces a common framework. The level of adoption of analytical tools can be assessed in one or more ways. This assessment in this particular case is made in two ways: first, questionnaires collect information regarding four attributes; and secondly, interviews are related to three additional attributes.

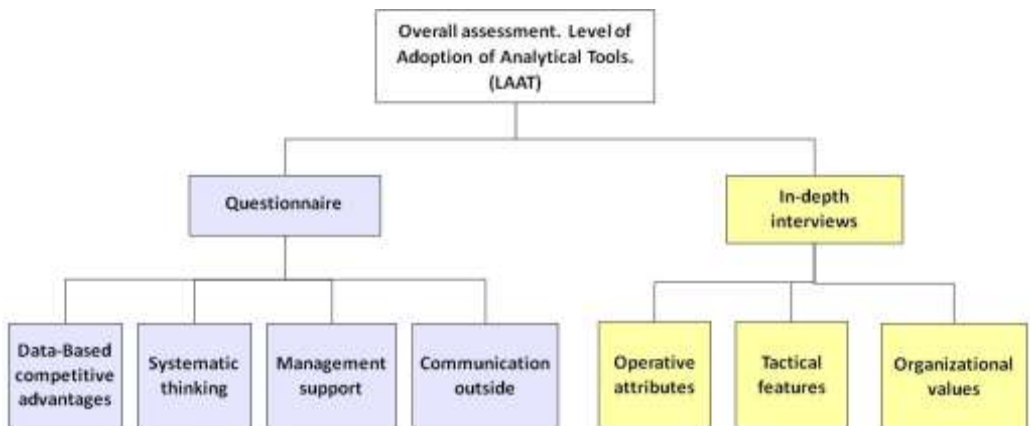

**Figure1.** A common framework for assessing the level of adoption of analytical tools.

On the left side of figure 1 is the first attribute of Data-Based competitive advantages. This refers to practices and actions that are implemented as a data analysis by organisations in order to create competitive advantages. In the second place, we have systemic thinking which measures the degree of systemic vision of the organisation. The third attribute refers to the degree of management support for analytical projects, and the last attribute relates to questionnaires, i.e. the degree of communication with outside actors. Against this, on the right side of figure 1, the first attribute with respect to interviews is organisational values', understood as a set of beliefs that specify universal expectations and modes of behaviour preferred by the organisation as they relate to data analysis and exploitation (Schwartz, 1994). This is followed by 'tactical features', referring to immediate or short-term actions that are less important than organisational values, but indispensable in terms of expanding the use of analytical tools. Well-aligned tactical features are expected to create a link between operative attributes and values. The last attribute that depends on interviews is the so-called 'operative attributes'. These are critical; this is where the "*rubber hits the road*" in terms of expanding the level of adoption of analytical tools through statistical methods, procedures and other analytical tools.

### 3.1 Written questionnaire

A questionnaire was designed to investigate the level of adoption of analytical tools. Several statistical tests were run in order to guarantee its reliability and validity, among them the interclass correlation coefficient (ICC) proposed by Fleiss (1971) and Shrout & Fleiss (1979), a measure of reliability on the scale proposed by Cronbach (1951) and the Kappa measure of agreement introduced by Cohen (1960). The results from all tests were satisfactory. A confirmatory factor analysis (CFA) later provided a quantitative foundation for the conceptual model. Table 2 shows the structure of the questionnaire.

**Table2**. Questionnaire structure.

| Section | Number of items |
|---|---|
| Categorical questions | 3 |
| Data Based Competitive Advantage | 5 |
| Management Support Data Analysis | 6 |
| Systemic Thinking | 5 |
| Communication outside the company | 1 |
| **Total** | **20** |

A total of 255 companies provided us with information related to their analytical capabilities. Questions were presented on a five-level scale, all related to data analysis best practices. The instrument was addressed to the information technologies manager, quality manager or managing director, with a request to be redirected to the proper person when necessary. On the cover stated, we openly stated our willingness to share the study conclusions with anyone who wanted them. Since the original instrument had 20 questions, only 17 were suitable for MCDA. Those responses, which were based on an ordinal scale with five assessment grades, are subjective in nature. The scale used is represented as follows:

$$H_1 = \{\ 'H_{1,1} - Worst', 'H_{1,2} - Poor', \\ \quad 'H_{1,3} - Average', 'H_{1,4} - Good\ ', \\ \quad 'H_{1,5} - Excellent'\} \tag{1}$$

According to the expression (1), a manager may choose one of the grades in order to assess the level of adoption of analytical tools in his/her organisation. Considering that $K$ responders participated in our study and $k_{l,n}$ of them selected a grade $H_{1,n}$ for assessing the organisation in alternative $A_l$, then the degree of belief $\beta_{1,n}^l$ to which an organisation assessed by the whole group of managers to the grade $H_{1,n}$ on the alternative $A_l$ is given as follows:

$$\beta_{1,n}^l = \frac{k_{l,n}}{K} \tag{2}$$

The rating that evaluates an organisation for alternative $A_l$ by the whole group of managers who answered our survey is expressed as:

$$S(A_l) = \{(H_{1,1}, \beta_{1,1}^l), (H_{1,2}, \beta_{1,2}^l), (H_{1,3}, \beta_{1,3}^l), (H_{1,4}, \beta_{1,4}^l), (H_{1,5}, \beta_{1,5}^l), (H_1, \beta_{H_1}^l)\} \tag{3}$$

Regarding expression (3), $0 \leq \beta_{1,n}^l \leq 1$. In addition, $\sum_{n=1}^{5} \beta_{1,n}^l \leq 1$ and $\beta_{H_1}^l = 1 - \sum_{n=1}^{5} \beta_{1,n}^l$ provide a measure for responders who did not make any assessment of alternative $A_l$. That is, $\beta_{H_1}^l$ represents the amount of missing information or the degree of ignorance for alternative $A_l$. According to Yang (2001) and Yang et al (2011), expression (3) adequately records the assessment information gathered and maintains the diversity of each questionnaire, and therefore, generates suitable information for further decision analysis. Moreover, and considering that our data comes from a survey, the mean calculated for the distributed assessment is relevant as a simpler performance indicator. If $u(H_{1,n})$ is the utility given to $H_{1,n}$ and there is no missing information, then $\beta_{H_1}^l = 0$, the mean for distribution (3) is given by:

$$u\big(S(A_l)\big) = \sum_{n=1}^{5} \beta_{1,n}^l \, u(H_{1,n}) \tag{4}$$

Formula (4) provides relevant information about the level of adoption of analytical tools for alternative $A_l$. For instance, if an organisation is given a high mean for any particular alternative, then that organisation should work to maintain the achieved strength. On the other hand, if the organisation obtains a low mean on any given alternative, then that alternative should be given a high priority so that the organisation can overcome this weakness. In short, expressions (1) to (4) can be applied to our data in order to collect relevant evidence regarding the level of adoption of analytical tools, including distributed assessments for each company. Comparisons can later be made between different companies on any given attribute.

### 3.2 Interviews

Interviews were performed to investigate the soft and unstructured features of the level of adoption of analytical tools. Although these interviews were unstructured, the script allowed us to maintain a general guideline during the course of each interview. The script and interviews were designed following the laddering methodology proposed by Reynolds & Gutman (1988). The term "*laddering*" refers to a one-on-one interviewing technique which is applied to understand how individuals transform the attributes of any given concept or idea into meaningful associations with respect to self by following the Means-End theory. In this research, we focused on investigating scales, but a detailed explanation of both laddering technique and Means-End theory can be found in Herrmann et al (2000), Reynolds & Gutman (1984) and Reynolds & Gutman (1988). The core idea behind the laddering technique basically involves eliciting elements in a sequential order from bottom to top. On the bottom, we find the most concrete (or least abstract) elements, while the top is composed of the most abstract elements. Three levels of abstraction were used following an order of "*attributes*" → "*tactical features*→ "*values*". in this respect, Deming (2000) emphasises the relevance of the values as a key element for expanding the adoption of analytical tools. Likewise and according to Anderson-Cook et. al (2012) (,) tactical features are the link between operative attributes and values. Based on this, a three-level scale was defined as follows:

$$H_2 = \{'H_{2,1} - Minimal\ impact', 'H_{2,2} - Average'$$

$$'H_{2,3} - Excellent'\} \tag{5}$$

Similar to the process applied to questionnaires as expressed in (3), the distributed assessment for the interviews in alternative $A_l$, given by:

$$S(A_l) = \left\{ (H_{2,1}, \beta_{2,1}^l), (H_{2,2}, \beta_{2,2}^l), (H_{2,3}, \beta_{2,3}^l), (H_2, \beta_{H_2}^l) \right\} \tag{6}$$

Where $\beta_{2,n}^l$, $n=1, 2, 3$ is calculated in the same way as expression (2). In addition, $\beta_{H_2}^l$ is a measure of ignorance, $0 \leq \beta_{2,n}^l \leq 1$ and $\sum_{n=1}^3 \beta_{2,n}^l \leq 1$. Likewise, $u(H_{2,n})$ is the utility assigned to $H_{2,n}$. If we assume that this is a complete distribution, so that $\beta_{H_2}^l = 0$, then the mean value is given by:

$$u(S(A_l)) = \sum_{n=1}^3 \beta_{2,n}^l \, u(H_{2,n}) \tag{7}$$

Expression (7) can be applied to decide if a criterion should be given high priority or to compare several organisations. For instance, if an organisation receives a higher accumulated degree of belief to the top grade, ($H_{2,3}$ in (6)) then this criterion should be given high priority in order to maintain the organisation's strengths. On the other hand, if an organisation receives the lowest accumulated degree of belief to the bottom grade ($H_{2,1}$ in (6)), then this criterion should be given high priority in order to improve the organisation's weakness.

At this point, scales (1) and (5) have to be transformed into a common framework in order to obtain a richer assessment on the level of adoption of analytical tools. This enriched assessment will provide more accurate information about how the adoption of analytical tools can be expanded. The following paragraphs will investigate a set of rule-based techniques to transform the original data into a common framework.

### 3.3 Common framework

The challenge, at this point, is how to aggregate two sources of information and investigate them under a single framework, while preventing the loss or skewing of any information. Yang et al (2011), Yang (2001) and Liu et. al (2008) demonstrated that expert judgments are routinely used in industry for interpreting survey data. Belton & Stewart (2002) highlighted the importance of the judge's expertise in enriching distribution assessments. The proposed scale, by gathering evidence from expert knowledge, should preserve original information from questionnaires and interviews while it is understandable and easy to use. The evidence gathered should likewise provide a set of common sense rules that can be used during the transformation process in a flexible form. Considering the above, a five-level monotonic scale is provided in the following way:

$H_3 = \{ \, 'H_1 - Analytics\ Ignorance', \, 'H_2 - Analytics\ focused',$

$\qquad 'H_3 - Analytical\ aspirations', 'H_4 - Systemic\ analytics',$

$\qquad 'H_5 - Analytics\ as\ competitive\ advantages' \} \tag{8}$

A complete and specific definition of the scale, including each of its five levels, was carried out while this research was performed. This is part of the operative definition of variables that was previously done in order to gather richer evidence. The distributed assessment of alternative $A_l$ including both questionnaires and interviews is given in expression (9).

$$S(A_l) = \left\{ (H_1, \beta_{1,}^l), (H_2, \beta_{,2}^l), (H_3, \beta_{,3}^l), (H_4, \beta_{,4}^l), (H_5, \beta_{,5}^l), (H, \beta_H^l) \right\}$$ (9)

Expressions (8) and (9) represent the common framework on which data from questionnaires and interviews are transformed. Moreover, the possibility that responders might select two or more assessment grades should be considered in the new common framework. For instance, if a responder ticks only one grade, then this is equivalent to assigning a belief of 100% to that grade and 0% to all others. The evaluation rating for each criterion is stated as follows:

$$\beta_{l,n} = \frac{1}{K} \sum_{j=1}^{K} \beta_{l,n,j}$$ (10)

Expression (10), $\beta_{l,n}$ represents the mean degree of belief given to the assessment $n$ on statement $l$-th. $K$ is the total number of respondents and $\beta_{l,n,l}$ is the degree of belief provided by the respondent $j$ to the assessment $n$. Furthermore, $\sum_{n=1}^{N} \beta_{l,n,j} \leq 1$ will be equal to 1.0 when the responder provides complete information; otherwise, this will be less than 1.0. This flexibility allows us to capture more accurate information and prevents a loss of significance.

**Table3**. Example of the new framework for assessment grades

| Assessment grade | Worst | Poor | Average | Good | Excellent | Unknown |
|---|---|---|---|---|---|---|
| Survey statement. "*Senior managers encourage data analysis for decision making*" | | | | | | |
| Belief of degree | $\beta_{l,1,j}$ | $\beta_{l,2,j}$ | $\beta_{l,3,j}$ | $\beta_{l,4,j}$ | $\beta_{l,5,j}$ | $\beta_{l,H,j}$ |

As mentioned before, the responder is not forced to tick only one grade. For instance, an individual manager may assess the statement "*senior managers encourage data analysis for better decision making*" as 50%="*average*", 30%="*good*" and 20%="*excellent*". Note that three grades were selected whose sum is equal to 1.0. With our common framework presented, now we must offer a set of rule-based techniques in order to transform the scales investigated.

### 3.4 Qualitative transformation for questionnaires

The scale used for the questionnaires can be almost directly transformed into the new common scale. That is, considering that both scales have five grades with the underlying logic that "*higher is better*", the transformation is easier to implement. The following equivalence of rules is proposed.

$'H_{1,1} - Worst'$ $\rightarrow$ $'H_1 - Analytics\ Ignorance'$

$'H_{1,2} - Poor',$ $\rightarrow$ $'H_2 - Analytics\ focused'$

$'H_{1,3} - Average'$ $\rightarrow$ $'H_3 - Analytical\ aspirations'$

$'H_{1,4} - Good\ '$ $\rightarrow$ $'H_4 - Systemic\ analytics'$

$'H_{1,5}$ $\rightarrow$ $'H_5 - Analytics\ as\ competitive\ advantages'$
$- Excellent'$

For purposes of this research, the symbol '$\rightarrow$' means '*is equivalent*" in terms of utility. The implementation of these rules does not imply that there is any change in the utilities. For instance, if $u(H_n)$ is defined as the utility of $H_n$, then $u(H_{1,1}) = u(H_1)$, $u(H_{1,2}) = u(H_2)$, $u(H_{1,3}) = u(H_3)$, $u(H_{1,4}) = u(H_4)$ and $u(H_{1,5}) = u(H_5)$. It is important to mention that we assume that the grades are evenly distributed in the assessment space: $H_1$ has the lowest utility while $H_5$ is associated with the highest.

According to these rules for transformation, the grades from the questionnaires can be transformed into the framework presented in subsection 2.3. That is, the distributed assessment presented in (3) is converted to its equivalent shown in (9).

### 3.5 Qualitative transformation of interviews

On the other hand, data from interviews is based on a scale of three levels. This can be expanded to five levels, representing an additional degree of complexity. The scale for the interviews is given the logical order "*higher is better*", and anchoring points are not required to carry out the transformation. The following rules of equivalence are proposed for interviews.

$'H_{2,1} - Minimal\ impact'$ $\rightarrow$ $'H_1 - Analytics\ Ignorance'$

$'H_{2,2} - Average\ impact'$ $\rightarrow$ $0.25'H_2 - Analytics\ focused'+$
$0.50'H_3 - Analytical\ aspirations'+$
$0.25'H_4 - Systemic\ analytics'$

$'H_{2,3} - Highest\ impact'$ $\rightarrow$ $'H_5 - Analytics\ as\ competitive\ advantages'$

The introduction of the proposed rules implies a change in the utilities. More specifically, we see that $u(H_{2,1}) = u(H_1)$, $u(H_{2,2}) = 0.25u(H_2) + 0.50u(H_3) + 0.25u(H_4)$ and $u(H_{2,3}) = u(H_5)$. There is also an assumption that the grades are evenly distributed. In this way, the distributed assessment presented in (6) is converted to its equivalent shown in (9). The next section gives a numerical example.

### 4. Numerical example

A six-step process is proposed to implement the described methodology. It begins with preparing the data for the analysis and finishes with the calculations (see figure 2).

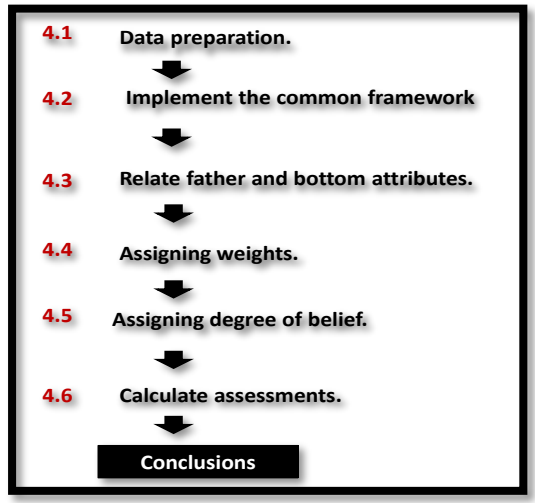

**Figure2**. Implementation procedure.

## 4.1 Data preparation

To begin several tasks must be completed in order to guarantee the data's suitability for the analysis. Some features must be observed for the dataset prior to the analysis. For instance, its format should allow it to be easily manipulated in order to establish common equivalences for data coming from different sources. Considering that poor quality data inevitably leads to incorrect conclusions, such tasks as purging and cleaning are our major concern on this step.

## 4.2 Implement the common framework

Our instrument is composed of 17 items, each on a five-level scale. These items are also classified according to four dimensions or parent attributes. On the other hand, interviews are composed of 3 parent-attributes and 25 bottom-level attributes, each on a three-level scale. The challenge is to aggregate both datasets into a common framework (see figure 1). Tables 4 and 5 present the structures and the related grades values for both questionnaires and interviews, respectively. The utilities assigned by us for each grade on both scales are based on the research conducted by Tallon, Kraemer & Gurbaxani (2000), who introduced a seven-level scale to measure the value of the business in a sample of 304 executives worldwide. Powell & Dent-Micallef (1997) also proposed a five-level scale to measure the degree of contribution of information technology to competitive advantages, and Gardner (2004) documented a five-level scale to measure the degree of maturity of a process. We assigned the degree of utility for each of the scales by carefully reviewing the scales, as seen in tables 4 and 5.

**Table4**. Structure and grades-values for questionnaires.

| Model summary | Grades for questionnaires |
|---|---|
| *Number of parent attributes: 4* | $u(H_{1,1}):= u\ (worst) =0.00$ |
| | $u(H_{1,2}):= u\ (poor) = 0.25$ |
| *Number of bottom atributes: 17* | $u(H_{1,3}):= u\ (average) =0.50$ |
| *Selected method for relating parent and bottom attributes:* | $u(H_{1,4}):= u\ (good) = 0.75$ |
| *RULE-BASED APPROACH* (Yang 2001) | $u(H_{1,5}):= u\ (best) = 1.00$ |

**Table5**. Structure and grades-values for interviews.

| Model summary | Grades for interviews |
|---|---|
| *Number of parent attributes: 3* | $u(H_{2,1}) := u$ (*Minimal*) =0.00 |
| *Number of bottom attributes: 25* | $u(H_{2,2}) := u$ (*Average*) = 0.50 |
| *Selected method for relating parent and bottom attributes: RULE-BASED APPROACH* (Yang 2001) | $u(H_{2,3}) := u$ (*Excellent*) =1.00 |

At this point, a common framework is proposed. The different scales, which were used to collect data from different sources, have been transformed to equivalent values by implementing rules. This problem is discussed in the next subsection.

### 4.3 Relate father and bottom attributes

According to Yang (2001), a quantitative relationship must be established between parent and bottom attributes. For instance, if the bottom attribute 'Communication and trust' is assessed on the basis of a 3-level scale, then it must be related to the father attribute given to it on a 5-level scale. This will require that we make transformations on 4 and 3 attributes for questionnaires and interviews, respectively, in order to complete the overall assessment of the model. Yang (2001) discusses two ways to convert bottom assessments to fathers, the first based on rules and the second on utilities. Due to limitations of space, we will not discuss here the properties of each transformation method. We selected the rule based method considering the advantages widely discussed by Xu, McCarthy & Yang (2006) and Yang (2001). The following is an illustrative example which will help to explain how this method works.

Consider the case of when a particular decision maker (DM) ticks the grade "*minimal impact*" on the attribute "*communication and trust*". A rule is generated to relate its grade with the overall performance in the following way: <u>IF</u> "*communication and trust*" <u>IS</u> "minimal", <u>THEN</u> overall performance is "*analytical ignorance*". Similarly, <u>IF</u> "*communication and trust*" <u>IS</u> "*average*", <u>THEN</u> overall performance is "*Analytical aspirations*". The complete set of rules for transforming this attribute is shown in table 6.

Table6. Relating COMMUNICATION AND TRUST to its father attribute.

| If COMMUNICATION AND TRUST is ***Minimal*** =0.00 | Then Overall Performance is ***Analytical Ignorance***=100% |
|---|---|
| If COMMUNICATION AND TRUST is ***Minimal***=0.125 & ***Average***=0.125 | Then Overall Performance is ***Analytical Focused***=100% |
| If COMMUNICATION AND TRUST is ***Average***=0.50 | Then Overall Performance is ***Analytical Aspirations***=100% |
| If COMMUNICATION AND TRUST is ***Average***=0.375 & ***Excellent***=0.375 | Then Overall Performance is ***Analytics as System***=100% |
| If COMMUNICATION AND TRUST is ***Excellent***=1.00 | Then Overall Performance is ***Analytics as Comp. Advantages***=100% |

A total of 210 rules were similarly generated in order to complete our model as shown in table 6, 85 of which were implemented for the 17 items of the questionnaire and 125 for the 25 interview attributes. Appendix A gives the complete list of bottom attributes.

### 4.4 Assigning weights

The weight of an attribute is its relative importance with respect to the rest of attributes in the model. That is, different features may have different importance, and this should be reflected in the model. For instance, according to Amabile et al (1996), if an organisation is willing to increase the use of analytical tools for better decision making, values might be more important than operative attributes. Consequently, organisational values should have a larger weight in the model. Here, we are implementing the weight assignment process proposed by Xu, McCarthy & Yang (2006). First, we calculate the frequency for each attribute by including all of its responses. The higher the frequency of a bottom-level attribute, the larger the weight it should be given in the model. Two of three types of weights were assigned in this way; the last was assigned based on the conclusions of literature. The lowest levels are weights assigned to the bottom-level attributes, 17 for questionnaires and 25 interviews.

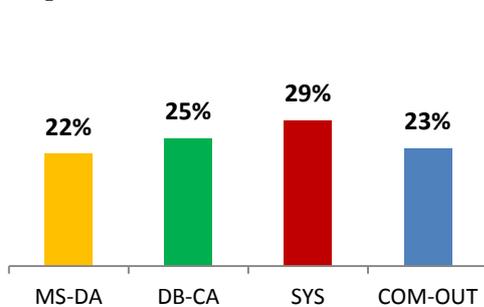

**Figure3**. Normalised weights for questionnaires.

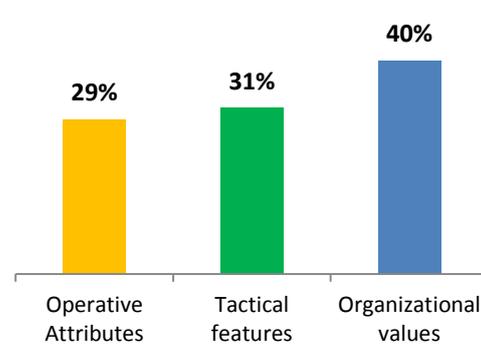

**Figure4**. Normalised weights for interviews.

Previous calculations clearly show that the Dempster rule proved to be commutative and associative – evidence that can be combined in any order. Although weights here were obtained by normalising responses from questionnaires and interviews, note that the combination of evidence can be carried out pairwise as well.

### 4.5 Assigning degree of belief

While the mean was calculated for each questionnaire attributed as $\bar{x} = \frac{1}{n}\sum_{i=1}^{n} x_i$, interviews were treated differently since they were given in terms of cumulative frequencies. The formula $y = 6.7x$ was applied to transform operative attributes; and similarly, tactical attributes and organisational values were transformed with $y = 7.3x$ and $= 4.3x$, respectively. Three of these were transformed to a three level scale (see appendix A).

At this point, both sources of data were ready for assignment of the degree of belief, meaning their conversion to a five level scale. For instance, if the mean of the attribute "*DB-CA5 In your company, is there a work environment that encourages the use of analytical techniques and data* *analysis?"* was $\bar{x}$ =2.236, then this value had to be transformed to the vector: {'*Worst*'=0.0, '*Poor*'=0.64, '*Average*'=0.36, '*Good*'=0.0, '*Excellent*'=0.0}. Similarly, if the mean value of the interview attribute "*Data online supports business decisions*" was $\bar{x}$ =2.25, then it should be transformed to the vector: {'*Minimal*'=0.0, '*Average*'=0.75, '*Excellent*'=0.25}. A total of 42 vectors, of which 17 were for questionnaires and 25 for interviews, were yielded. Bellow, the general formulation that summarises the performed transformations is shown.

$$g(\bar{x})_i = \begin{cases} 0 & \text{if} & \lceil \bar{x} \rceil > i > \lfloor \bar{x} \rfloor \\ \bar{x} - \lfloor \bar{x} \rfloor & \text{if} & i = \lceil \bar{x} \rceil \\ \lceil \bar{x} \rceil - \bar{x} & \text{if} & i = \lfloor \bar{x} \rfloor \end{cases} \qquad (i=1...5) \qquad (11)$$

According to expression (11), $g(\bar{x})_i$ is equal to 0 if $\lceil \bar{x} \rceil > i > \lfloor \bar{x} \rfloor$. A second possibility is $g(\bar{x})_i = \bar{x} - \lfloor \bar{x} \rfloor$ when $i = \lceil \bar{x} \rceil$. Finally $g(\bar{x})_i = \lceil \bar{x} \rceil - \bar{x}$, in the case $i = \lfloor \bar{x} \rfloor$.

Similar expression was applied on interviews, as seen below.

$$h(\bar{x})_j = \begin{cases} 0 & \text{if} & \lceil \bar{x} \rceil > i > \lfloor \bar{x} \rfloor \\ \bar{x} - \lfloor \bar{x} \rfloor & \text{if} & i = \lceil \bar{x} \rceil \\ \lceil \bar{x} \rceil - \bar{x} & \text{if} & i = \lfloor \bar{x} \rfloor \end{cases} \qquad (j=1...3) \qquad (12)$$

On (12) $h(\bar{x})_j$ will be equal to 0 if $\lceil \bar{x} \rceil > i > \lfloor \bar{x} \rfloor$; and in the same way $h(\bar{x})_j = \bar{x} - \lfloor \bar{x} \rfloor$ when $i = \lceil \bar{x} \rceil$. Finally $h(\bar{x})_j = \lceil \bar{x} \rceil - \bar{x}$, in the case $i = \lfloor \bar{x} \rfloor$. Numerical results of these on appendix A are provided.

### 4.6 Calculate assessments

The 210 rules that were previously mentioned were implemented in order to calculate the distributed assessments. Doing this on a spreadsheet might be difficult and time consuming due to the complexity of the calculations. The Intelligent Decisions Systems (IDS) package developed by Yang & Xu (2005) was used for these purposes. This software, which was designed based on the evidential reasoning approach, allowed us to obtain the distributed assessments for each type of company (detailed information about the package can be reached on http://www.e-ids.co.uk/).

Although a total of 168 distributed assessments were obtained on the IDS, only two of them are discussed here due to space limitations. Figure 5A illustrates the assessment for the attribute "*Management support on data analysis*", which also follows the structure presented in formulation (3) with a 5-level scale. Figure 5B presents the distribution assessment for the organisational value "*Communication and trust*". Note that it follows the structure presented for formulation (6) with a 3-level scale. Although these assessments were calculated for overall performance of the model, it is feasible to obtain a total of 42 distributed assessments for each type of company. This means that the complete model is composed of 168 different distributed assessments. This allows us to preserve richer information regarding each type of company in order to rank them or provide more accurate evidence which will allow us to make more precise decisions related with the expansion on the use of analytical tools.

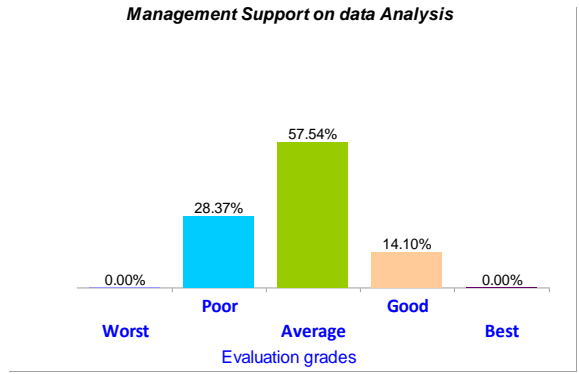

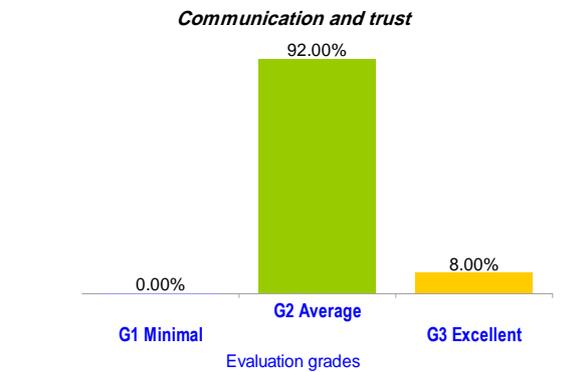

**Figure 5A**. Distributed assessment with 5 level scale

**Figure 5B**. Distributed assessment with 3 level scale

At this point, we have obtained the assessments; now we must fully implement the set of rules in order to have our complete model. For instance, if the distribution assessments presented in figures 5A and 5B are on a 5-level and 3-level scale, respectively, then they should be transformed to the common scale presented in formulation (8). Finally, a unique distributed assessment for the common framework which fits the structure of the equation (9) should be calculated for each type of organisation.

The IDS software was applied to aggregate lower level attributes first, and then higher level attributes. Consider the following cases as illustrative explanation: in the case of questionnaires, criteria $C(1,1,1)$ to $C(1,1,11)$ were aggregated in the higher level criterion $C(1,1)$. In the case of the interviews, the aggregation of attributes $C(2,1,1)$ to $C(2,1,11)$ resulted in the higher level attribute $C(2,1)$. The last step of the process consisted of aggregating five pairs of distribution assessments as follows: (1,1) with (2,1); (1,2) with (2,2); (1,3) with (2,3); (1,4) with (2,4); and finally (1,5) with (2,5). The distributed assessments for the whole model were obtained as follows:

$$S(Micro)$$
$$= \{(H_1, 0.0979), (H_2, 0.2340), (H_3, 0.2123), (H_4, 0.1201), (H_5, 0.3357)\}$$
$$S(Small)$$
$$= \{(H_1, 0.0905), (H_2, 0.1423), (H_3, 0.3569), (H_4, 0.0804), (H_5, 0.3299)\}$$
$$S(Medium)$$
$$= \{(H_1, 0.0896), (H_2, 0.0850), (H_3, 0.3873), (H_4, 0.1074), (H_5, 0.3308)\}$$
$$S(Large)$$
$$= \{(H_1, 0.0892), (H_2, 0.1038), (H_3, 0.3588), (H_4, 0.1204), (H_5, 0.3277)\}$$

Based on these distributed assessments, it is clear that large and medium size organisations are more analytically oriented than micro and small size companies. Nevertheless, it is impossible to directly ascertain which type of organisation is the most analytically oriented. Based on a slight difference, either large or medium size companies could be the most analytically oriented. In order to more accurately prioritise the organisations investigated, we prepare a chart by first quantifying the assessments grades and then calculating the mean scores by including the 42 attributes of the whole model. This is done for each type of organisations. The next figure presents the ranking, based on their overall performance.

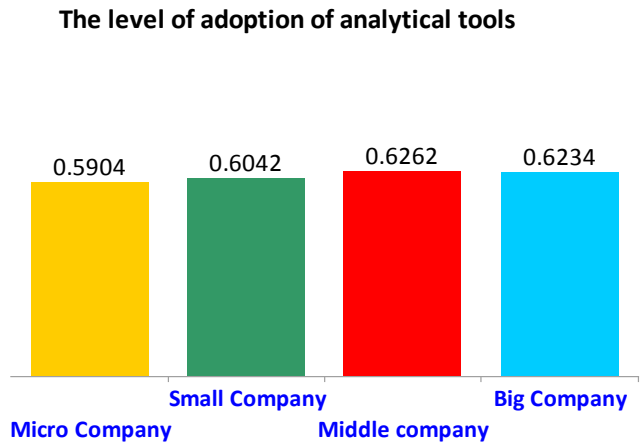

**The level of adoption of analytical tools**

**Figure6**. Ranking of companies according their analytical capabilities.

According to figure 6, the following ranking order for the four types of companies is obtained: Medium ≻ Large ≻ Small ≻ Micro. For purposes of this research, the symbol " ≻ " represents "*is more analytically oriented than*". The foregoing ranking was calculated assuming that interviews have 60% of the weight and questionnaires the remaining 40%. This is coherent with discussions on the literature reviewed, which showed that organisational values have a deeper impact on the adoption of analytical tools.

## 5. Concluding remarks

This paper investigates a methodology that can be used to model measures for the level of adoption of analytical tools. Here, are analysed two datasets that are different in nature and collected from distinct sources. The principal question is how to integrate data from different sources but related to the same research topic into a common framework, while preventing information from being lost or skewed. The knowledge and judgments of experts should also be systematically and consistently incorporated in the proposed common framework in order to yield richer evidence for making more accurate decisions regarding the expansion of the adoption of analytical tools. In order to achieve the above, a set of novel and pragmatic transformation rules were investigated and later implemented in an example.

When doing research, it is common to cope with information that comes from different sources but is related to the same topic. For instance, consider the case where data that is received from surveys, research literature, interviews, web pages and other sources needs to be handled and integrated in order to draw unique conclusions. The richness and diversity of the data should also remain unaltered and retain its original identity. Here, the challenge is to aggregate this data in a systemic, pragmatic and consistent way. The methodology explained here provides researchers with a practical, flexible and consistent tool that can be used to tackle the challenge of aggregating multiple source data. A numerical example is provided in order to make the methodology easier to understand. Although this case investigates the aggregation of two datasets, the methodology can be replicated for three or more different types of data, regardless of whether the data is quantitative or qualitative. A six-step process is provided in this respect as a guideline for practitioners and researchers who are interested in tackling problems with multiple-source data. A specialised software is introduced to assist them in coping with the complexity of the calculations. In this way, anyone interested can easily design their own models, aggregate multisource data and prioritize alternatives.

The primary purpose of this paper is to demonstrate the usefulness of this methodology, discuss its most relevant features and provide a numerical example. It is evident that important findings related to the adoption of analytical tools were not fully explained. Due to the richness and diversity of the information obtained, it was not possible to provide a deep and complete discussion here.  A careful investigation of the information generated in this model will help to identify the most important key drivers that yield the expansion on the use of analytical tools on organisations. This investigation is subject to further research.

## 6.  References

Amabile, T. M., Conti, R., Coon, H., Lazenby, J., & Herron, M. (1996). Assessing the work environment for creativity. *Academy of Management Journal, 39*(5), 1154-1184.

Anderson-Cook, C. M., Lu, L., Clark, G., DeHart, S. P., Hoerl, R., Jones, B., et al. (2012).Statistical Engineering-Forming the Foundations. *Quality Engineering, 24*(2), 110-132.

Barahona, I., & Riba, A. (2012). Applied Statistics on Business at Spain: A Case of Statistical Engineering. In A. S. Association. (Ed.). In *Joint Statistical Meetings* [Contributed]. San Diego, CA:(American Statistical Association.).

Belton, V., & Stewart, T. (2002). *Multiple Criteria Decision Analysis: An Integrated Approach*: Springer US.

Brans, J.-P., Vincke, P., & Mareschal, B. (1986). How to select and how to rank projects: The PROMETHEE method. *European journal of operational research, 24*(2), 228-238.

Cohen, J. (1960). A COEFFICIENT OF AGREEMENT FOR NOMINAL SCALES. *Educational and Psychological Measurement, 20*(1), 37-46.

Cronbach, L. J. (1951). COEFFICIENT ALPHA AND THE INTERNAL STRUCTURE OF TESTS. *Psychometrika, 16*(3), 297-334.

Davenport, T. H., Harris, J. G., & Morison, R. (2010). *Analytics at work: smarter decisions, better results*. Boston, MA: Harvard Business Press.

Deming, W. E. (2000). *Out of the Crisis*: MIT press.

Fleiss, J. L. (1971). MEASURING NOMINAL SCALE AGREEMENT AMONG MANY RATERS. *Psychological Bulletin, 76*(5), 378-382.

Gantz, J., & Reinsel, D. (2012). THE DIGITAL UNIVERSE IN 2020: Big Data, Bigger Digital Shadows, and Biggest Grow in the Far East. (EMC Corporation).

Gardner, R. (2004). *The Process-focused Organization: A Transition Strategy for Success*: ASQ Quality Press.

Hayashi, K. (2000). Multicriteria analysis for agricultural resource management: a critical survey and future perspectives. *European Journal of Operational Research, 122*(2), 486-500.

Herrmann, A., Huber, F., & Braunstein, C. (2000). Market-driven product and service design: Bridging the gap between customer needs, quality management, and customer satisfaction. *International Journal of Production Economics, 66*(1), 77-96.

Liu, X.-B., Zhou, M., Yang, J.-B., & Yang, S.-L. (2008). Assessment of strategic R&D projects for car manufacturers based on the evidential reasoning approach. *International Journal of Computational Intelligence Systems, 1*(1), 24-49.

Louviere, J. J., Hensher, D. A., & Swait, J. D. (2000). *Stated choice methods: analysis and applications*: Cambridge University Press.

Lynch, C. (2008). Big data: How do your data grow? *Nature, 455*(7209), 28-29.

Malczewski, J. (1999). *GIS and multicriteria decision analysis*: John Wiley & Sons.

Mendoza, G., & Martins, H. (2006). Multi-criteria decision analysis in natural resource management: a critical review of methods and new modelling paradigms. *Forest ecology and management, 230*(1), 1-22. Mendoza, G., & Prabhu, R. (2002). *Enhancing participatory planning of community-managed forest using problem structuring models and approaches: experiences from a case study*: Working Paper GAM–2003–1. Dep. Natural Resour. Environ. Sci., University of Illinois, Champaign.

Mousseau, V., & Slowinski, R. (1998). Inferring an ELECTRE TRI model from assignment examples. *Journal of global optimization, 12*(2), 157-174.

Mousseau, V., Slowinski, R., & Zielniewicz, P. (2000).A user-oriented implementation of the ELECTRE-TRI method integrating preference elicitation support. *Computers & operations research, 27*(7), 757-777.

Pohekar, S., & Ramachandran, M. (2004).Application of multi-criteria decision making to sustainable energy planning—a review. *Renewable and Sustainable Energy Reviews, 8*(4), 365-381.

Powell, T. C., & DentMicallef, A. (1997). Information technology as competitive advantage: The role of human, business, and technology resources. *Strategic Management Journal, 18*(5), 375-405.

*Predictive analytics* (2014). Retrieved 01 of July of 2014, 2014, from http://www.answers.com/topic/predictive-analytics

Pukkala, T. (2002). *Multi-objective forest planning*: Kluwer academic publishers.

Reynolds, T. J., & Gutman, J. (1984). Laddering: Extending the Repertory Grid Methodology to Construct Attribute-Consequence-Value Hierarchies. In *in Personal Values and Consumer*. (Books).

Reynolds, T. J., & Gutman, J. (1988). LADDERING THEORY, METHOD, ANALYSIS, AND INTERPRETATION. *Journal of Advertising Research, 28*(1), 11-31.

Rousseau, D. M. (2006).Is there such a thing as "evidence-based management"? *Academy of Management Review, 31*(2), 256-269.

Roy, B. (1968). Classementetchoix en présence de points de vue multiples. *RAIRO-Operations Research-RechercheOpérationnelle, 2*(V1), 57-75.

Saaty, T. L. (1990). How to make a decision: the analytic hierarchy process. *European journal of operational research, 48*(1), 9-26.

Saaty, T. L. (1991). *Rank and the Controversy About the Axioms of Utility Theory--A Comparison of AHP and MAUT.* Paper presented at the Proceedings of the 2nd International Symposium of The Analytic Hierarchy Process.

Satty, T. L. (1980).The analytic hierarchy process. (New York: McGraw-Hill New York).

Scott, A. J. (2012). Moneyball: Message for Managers [Marketing paper]. Pennsylvania:(University of Pennsylvania ScholarlyCommons ).

Shrout, P. E., & Fleiss, J. L. (1979). INTRACLASS CORRELATIONS - USES IN ASSESSING RATER RELIABILITY. *Psychological Bulletin, 86*(2), 420-428.

Schwartz, S. H. (1994). Are there universal aspects in the structure and contents of human values?. Journal of social issues, 50(4), 19-45.

Tallon, P. P., Kraemer, K. L., & Gurbaxani, V. (2000). Executives' perceptions of the business value of information technology: A process-oriented approach. *Journal of Management Information Systems, 16*(4), 145-173.

Tzeng, G.-H., & Huang, J.-J. (2011). *Multiple attribute decision making: methods and applications*: CRC Press.

Xu, D.-L., McCarthy, G., & Yang, J.-B. (2006). Intelligent decision system and its application in business innovation self assessment. *Decision Support Systems, 42*(2), 664-673.

Yang, J., & Xu, D. (2005).The IDS multi-criteria assessor software. *Intelligent Decision System, Cheshire, UK*.

Yang, J.-B., & Singh, M. G. (1994). An evidential reasoning approach for multiple-attribute decision making with uncertainty. *Systems, Man and Cybernetics, IEEE Transactions on, 24*(1), 1-18.

Yang, J. B. (2001). Rule and utility based evidential reasoning approach for multiattribute decision analysis under uncertainties. *European Journal of Operational Research, 131*(1), 31-61.

Yang, J. B., Xu, D. L., Xie, X., & Maddulapalli, A. K. (2011). Multicriteria evidential reasoning decision modelling and analysis-prioritizing voices of customer. *Journal of the Operational Research Society, 62*(9), 1638-1654.

Yoon, K. P., & Hwang, C.-L. (1995). *Multiple attribute decision making: an introduction* (Vol. 104): Sage Publications.

**Appendix.**

**Bottom level attributes. Interviews.**

| | Count | Concept | Frequency | % relative | mean | minimal | average | excellent |
|---|---|---|---|---|---|---|---|---|
| **O-A** | 1 | Data is accessible and supports decisions | 20.00 | 0.20 | 3.00 | | | 1.00 |
| **O-A** | 2 | Data online supports decisions | 15.00 | 0.15 | 2.25 | | 0.75 | 0.25 |
| **O-A** | 3 | Goal setting | 13.00 | 0.13 | 1.95 | 0.05 | 0.95 | |
| **O-A** | 4 | Standardized procedures | 11.00 | 0.11 | 1.65 | 0.35 | 0.65 | |
| **O-A** | 5 | high skilled staff | 8.00 | 0.08 | 1.20 | 0.80 | 0.20 | |
| **O-A** | 6 | Enough support | 7.00 | 0.07 | 1.05 | 0.95 | 0.05 | |
| **O-A** | 7 | High tech | 6.00 | 0.06 | 0.90 | 1.00 | | |
| **O-A** | 8 | Communication with customers and suppliers | 5.00 | 0.05 | 0.75 | 1.00 | | |
| **O-A** | 9 | Creativity to propose new ideas | 5.00 | 0.05 | 0.75 | 1.00 | | |
| **O-A** | 10 | Information outside the organization | 5.00 | 0.05 | 0.75 | 1.00 | | |
| **O-A** | 11 | Market research | 5.00 | 0.05 | 0.75 | 1.00 | | |
| **T-A** | 1 | Improving data analysis | 22.00 | 0.22 | 3.00 | | | 1.00 |
| **T-A** | 2 | Improving results | 17.00 | 0.17 | 2.32 | | 0.68 | 0.32 |
| **T-A** | 3 | Financial benefits | 15.00 | 0.15 | 2.05 | | 0.95 | 0.05 |
| **T-A** | 4 | Staff efficiency and motivation | 12.00 | 0.12 | 1.64 | 0.36 | 0.64 | |
| **T-A** | 5 | Exceeding the customer expectations | 7.00 | 0.07 | 0.95 | 1.00 | | |
| **T-A** | 6 | Improving processes | 7.00 | 0.07 | 0.95 | 1.00 | | |
| **T-A** | 7 | Knowledge of data | 7.00 | 0.07 | 0.95 | 1.00 | | |
| **T-A** | 8 | Long term relationships with actors | 7.00 | 0.07 | 0.95 | 1.00 | | |
| **T-A** | 9 | Distinctive competence | 6.00 | 0.06 | 0.82 | 1.00 | | |
| **O-V** | 1 | Add value to stake holders | 13.00 | 0.24 | 3.00 | | | 1.00 |
| **O-V** | 2 | Serving the society | 12.00 | 0.22 | 2.77 | | 0.23 | 0.77 |
| **O-V** | 3 | Passion, Quality and Excellence | 11.00 | 0.20 | 2.54 | | 0.46 | 0.54 |
| **O-V** | 4 | Being a leader | 9.00 | 0.17 | 2.08 | | 0.92 | 0.08 |
| **O-V** | 5 | Communication and trust | 9.00 | 0.17 | 2.08 | | 0.92 | 0.08 |

| | |
|---|---|
| **O-V** | Organizational values |
| **T-A** | Tactical attributes |
| **O-A** | Operative attributes |

**Questionnaires.**
**Bottom level attributes. Questionnaires**

| Alternative | Code | N | Mean | StDv | Worst (1) | Poor (2) | Average (3) | Good (4) | Excellent (5) | CI-Low | CI-High | CI-Range |
|---|---|---|---|---|---|---|---|---|---|---|---|---|
| Micro Company | MS-DA1 | 86 | 1.76744 | 1.28039 | 0.25 | 0.75 | | | | 1.4968 | 2.0381 | 0.5412 |
| | MS-DA2 | 86 | 1.88372 | 1.13156 | 0.12 | 0.88 | | | | 1.6446 | 2.1229 | 0.4783 |
| | MS-DA3 | 86 | 2.06977 | 1.29061 | | 0.94 | 0.06 | | | 1.7970 | 2.3425 | 0.5455 |
| | MS-DA4 | 73 | 1.53425 | 1.10657 | 0.5 | 0.5 | | | | 1.2804 | 1.7881 | 0.5077 |
| | MS-DA5 | 73 | 2.10959 | 1.29702 | | 0.9 | 0.1 | | | 1.8121 | 2.4071 | 0.5951 |
| | MS-DA6 | 67 | 4.08955 | 0.99592 | | | | 0.92 | 0.08 | 3.8511 | 4.3280 | 0.4770 |
| | DB-CA1 | 73 | 2.87671 | 1.43319 | | 0.12 | 0.88 | | | 2.5479 | 3.2055 | 0.6575 |
| | DB-CA2 | 98 | 2.19388 | 1.32896 | | 0.81 | 0.19 | | | 1.9308 | 2.4570 | 0.5262 |
| | DB-CA3 | 98 | 2.82653 | 1.33953 | | 0.18 | 0.82 | | | 2.5613 | 3.0917 | 0.5304 |
| | DB-CA4 | 86 | 2.67442 | 1.22178 | | 0.33 | 0.67 | | | 2.4162 | 2.9326 | 0.5165 |
| | DB-CA5 | 86 | 2.06977 | 1.16610 | | 0.94 | 0.06 | | | 1.8233 | 2.3162 | 0.4929 |
| | SYS1 | 72 | 2.23611 | 1.25027 | | 0.77 | 0.23 | | | 1.9473 | 2.5249 | 0.5776 |
| | SYS2 | 67 | 2.88060 | 1.49263 | | 0.12 | 0.88 | | | 2.5232 | 3.2380 | 0.7148 |
| | SYS3 | 67 | 2.68657 | 1.35075 | | 0.32 | 0.68 | | | 2.3631 | 3.0100 | 0.6469 |
| | SYS4 | 66 | 3.80303 | 1.40570 | | | 0.20 | 0.80 | | 3.4639 | 4.1422 | 0.6783 |
| | SYS5 | 73 | 3.73973 | 1.21367 | | | 0.26 | 0.74 | | 3.4613 | 4.0181 | 0.5568 |
| | COM-OUT | 98 | 2.09184 | 1.32452 | | 0.91 | 0.09 | | | 1.8296 | 2.3541 | 0.5245 |
| Small Company | MS-DA1 | 66 | 2.04545 | 1.20807 | | 0.95 | 0.05 | | | 1.7540 | 2.3369 | 0.5829 |
| | MS-DA2 | 66 | 2.36364 | 1.19790 | | 0.64 | 0.36 | | | 2.0746 | 2.6526 | 0.5780 |
| | MS-DA3 | 66 | 2.30303 | 1.14985 | | 0.7 | 0.3 | | | 2.0256 | 2.5804 | 0.5548 |
| | MS-DA4 | 58 | 1.84483 | 1.18176 | 0.16 | 0.84 | | | | 1.5407 | 2.1490 | 0.6083 |
| | MS-DA5 | 58 | 2.32759 | 1.08236 | | 0.67 | 0.33 | | | 2.0490 | 2.6061 | 0.5571 |
| | MS-DA6 | 56 | 4.07143 | 1.04198 | | | | 0.93 | 0.07 | 3.7985 | 4.3443 | 0.5458 |
| | DB-CA1 | 58 | 3.17241 | 1.33952 | | | 0.83 | 0.17 | | 2.8277 | 3.5172 | 0.6895 |
| | DB-CA2 | 73 | 2.52055 | 1.20312 | | 0.48 | 0.52 | | | 2.2446 | 2.7965 | 0.5520 |
| | DB-CA3 | 73 | 2.95890 | 1.32747 | | 0.04 | 0.96 | | | 2.6544 | 3.2634 | 0.6090 |
| | DB-CA4 | 66 | 2.89394 | 1.21692 | | 0.11 | 0.89 | | | 2.6003 | 3.1875 | 0.5872 |
| | DB-CA5 | 66 | 2.42424 | 1.09630 | | 0.57 | 0.43 | | | 2.1598 | 2.6887 | 0.5290 |
| | SYS1 | 58 | 2.68966 | 1.18776 | | 0.31 | 0.69 | | | 2.3840 | 2.9953 | 0.6114 |
| | SYS2 | 56 | 3.01786 | 1.40766 | | | 0.983 | 0.017 | | 2.6492 | 3.3865 | 0.7374 |
| | SYS3 | 56 | 2.83929 | 1.21770 | | 0.16 | 0.84 | | | 2.5204 | 3.1582 | 0.6379 |
| | SYS4 | 56 | 3.32143 | 1.45361 | | | 0.68 | 0.32 | | 2.9407 | 3.7022 | 0.7614 |
| | SYS5 | 58 | 3.17241 | 1.12605 | | | 0.83 | 0.17 | | 2.8826 | 3.4622 | 0.5796 |
| | COM-OUT | 73 | 2.64384 | 1.37810 | | 0.34 | 0.64 | | | 2.3277 | 2.9600 | 0.6323 |

| Alternative | Code | N | Mean | StDv | Worst (1) | Poor (2) | Average (3) | Good (4) | Excellent (5) | CI-Low | CI-High | CI-Range |
|---|---|---|---|---|---|---|---|---|---|---|---|---|
| Middle company | MS-DA1 | 23 | 2.30435 | 1.29456 | | 0.70 | 0.30 | | | 1.7753 | 2.8334 | 1.0581 |
| | MS-DA2 | 23 | 2.30435 | 1.29456 | | 0.70 | 0.30 | | | 1.7753 | 2.8334 | 1.0581 |
| | MS-DA3 | 23 | 2.86957 | 1.42396 | | 0.13 | 0.87 | | | 2.2876 | 3.4515 | 1.1639 |
| | MS-DA4 | 21 | 1.85714 | 1.38873 | 0.14 | 0.86 | | | | 1.2632 | 2.4511 | 1.1879 |
| | MS-DA5 | 21 | 2.95238 | 1.35927 | | 0.05 | 0.95 | | | 2.3710 | 3.5338 | 1.1627 |
| | MS-DA6 | 20 | 4.20000 | 0.83351 | | | | 0.80 | 0.20 | 3.8347 | 4.5653 | 0.7306 |
| | DB-CA1 | 21 | 3.57143 | 1.39898 | | | 0.43 | 0.57 | | 2.9731 | 4.1698 | 1.1967 |
| | DB-CA2 | 25 | 3.00000 | 1.29099 | | | 1.00 | | | 2.4939 | 3.5061 | 1.0121 |
| | DB-CA3 | 25 | 2.92000 | 1.15181 | | 0.08 | 0.92 | | | 2.4685 | 3.3715 | 0.9030 |
| | DB-CA4 | 23 | 3.43478 | 1.03687 | | | 0.57 | 0.43 | | 3.0110 | 3.8585 | 0.8475 |
| | DB-CA5 | 23 | 3.00000 | 1.24316 | | | 1 | | | 2.4919 | 3.5081 | 1.0161 |
| | SYS1 | 21 | 2.85714 | 1.42428 | | 0.86 | 0.14 | | | 2.2480 | 3.4663 | 1.2183 |
| | SYS2 | 20 | 3.50000 | 1.31789 | | | 0.50 | 0.50 | | 2.9224 | 4.0776 | 1.1552 |
| | SYS3 | 19 | 3.26316 | 1.32674 | | | 0.74 | 0.26 | | 2.6666 | 3.8597 | 1.1931 |
| | SYS4 | 20 | 3.60000 | 1.09545 | | | 0.40 | 0.60 | | 3.1199 | 4.0801 | 0.9602 |
| | SYS5 | 21 | 2.85714 | 1.31475 | | 0.14 | 0.86 | | | 2.2948 | 3.4195 | 1.1247 |
| | COM-OUT | 25 | 2.92000 | 1.35154 | | 0.08 | 0.92 | | | 2.3902 | 3.4498 | 1.0596 |
| Big Company | MS-DA1 | 16 | 2.37500 | 1.25831 | | 0.63 | 0.37 | | | 1.7584 | 2.9916 | 1.2331 |
| | MS-DA2 | 16 | 2.62500 | 1.14746 | | 0.38 | 0.62 | | | 2.0627 | 3.1873 | 1.1245 |
| | MS-DA3 | 16 | 2.62500 | 1.36015 | | 0.38 | 0.62 | | | 1.9585 | 3.2915 | 1.3329 |
| | MS-DA4 | 13 | 2.15385 | 1.40512 | | 0.85 | 0.15 | | | 1.3900 | 2.9177 | 1.5277 |
| | MS-DA5 | 13 | 2.92308 | 1.18754 | | 0.077 | 0.923 | | | 2.2775 | 3.5686 | 1.2911 |
| | MS-DA6 | 13 | 3.53846 | 1.45002 | | | 0.46 | 0.54 | | 2.7502 | 4.3267 | 1.5765 |
| | DB-CA1 | 13 | 3.00000 | 1.29099 | | | 1.00 | | | 2.2982 | 3.7018 | 1.4036 |
| | DB-CA2 | 18 | 2.88889 | 1.36722 | | 0.112 | 0.888 | | | 2.2573 | 3.5205 | 1.2632 |
| | DB-CA3 | 18 | 3.33333 | 1.32842 | | | 0.67 | 0.33 | | 2.7196 | 3.9470 | 1.2274 |
| | DB-CA4 | 16 | 3.37500 | 1.36015 | | | 0.63 | 0.37 | | 2.7085 | 4.0415 | 1.3329 |
| | DB-CA5 | 16 | 3.06250 | 1.23659 | | | 0.94 | 0.06 | | 2.4566 | 3.6684 | 1.2119 |
| | SYS1 | 13 | 2.38462 | 1.26085 | | 0.616 | 0.384 | | | 1.6992 | 3.0700 | 1.3708 |
| | SYS2 | 13 | 3.23077 | 1.36344 | | 0.77 | 0.23 | | | 2.4896 | 3.9719 | 1.4824 |
| | SYS3 | 13 | 2.61538 | 1.26085 | | 0.385 | 0.615 | | | 1.9300 | 3.3008 | 1.3708 |
| | SYS4 | 13 | 3.53846 | 1.05003 | | | 0.46 | 0.54 | | 2.9677 | 4.1093 | 1.1416 |
| | SYS5 | 13 | 3.30769 | 1.25064 | | | 0.693 | 0.307 | | 2.6278 | 3.9875 | 1.3597 |
| | COM-OUT | 18 | 3.38889 | 1.61387 | | | 0.612 | 0.388 | | 2.6433 | 4.1345 | 1.4911 |